\documentclass[letterpaper,twocolumn,amsmath,amssymb,pre,aps,10pt]{revtex4-1}
\usepackage{graphicx}
\usepackage{dcolumn}
\usepackage{bm}
\usepackage{color}
\usepackage{subcaption} 

\usepackage{dcolumn}
\newcolumntype{d}[1]{D{.}{.}{#1}}

\newcommand{\rr}{\textbf{r}}

\newcommand{\derivation}[1]{} 

\begin{document}
\title{An efficient approach to approximating the pair distribution
  function of the inhomogeneous hard-sphere fluid}

\author{Paho Lurie-Gregg}
\author{Jeff B. Schulte}
\author{David Roundy}
\affiliation{Department of Physics, Oregon State University, Corvallis, OR 97331}

\begin{abstract}
We introduce an approximation for the pair distribution function of
the inhomogeneous hard sphere fluid. Our approximation makes use of
our recently published averaged pair distribution function at contact
which has been shown to accurately reproduce the averaged pair
distribution function at contact for inhomogeneous density
distributions. This approach achieves greater computational efficiency
than previous approaches by enabling the use of exclusively fixed-kernel
convolutions and thus allowing an implementation using fast Fourier
transforms. We compare results for our pair distribution approximation
with two previously published works and Monte-Carlo simulation,
showing favorable results.
\end{abstract}

\maketitle

\section{Introduction}

The standard approach in liquid state theory is to model a liquid as a
hard-sphere reference fluid with attractive interactions that are
treated perturbatively~\cite{hansen2006theory}.  Recent advances have
extended these perturbative approaches to inhomogeneous density
distributions---that is, liquid interfaces---through the use of
classical density functional theory (DFT), in which the grand free
energy is found by minimizing a free energy functional of the
density~\cite{ jain2007modified, gloor2007prediction,
  gross2009density,
  kahl2008modified,
  hughes2013classical,
  bryk2006density, clark2006developing,
  kahl2008modified, gross2009density,
  sundararaman2012computationally, marshall2013density}.  The
perturbation theory treatment of intermolecular interactions relies on
the pair distribution function of the reference fluid:
$g_{HS}^{(2)}(\rr_1,\rr_2)$.  Unlike the radial distribution function of a
homogeneous fluid, there does not currently exist a tractable form for
the pair distribution function of an inhomogeneous hard-sphere fluid,
suitable for use in constructing a density
functional~\cite{gloor2007prediction, jain2007modified}.

At its core, thermodynamic perturbation theory (TPT), sometimes referred to
as the high-temperature expansion, is an expansion of the
free energy in powers of a small parameter, which is the
product of a
pairwise attractive interaction with the inverse temperature $\beta$:
\begin{align}
  F &= F_0 + F_1 + \beta F_2 + \mathcal{O}(\beta^2)
\end{align}
where the terms $F_n$ are corrections to the free energy of order $n$
in the small interaction.  The first and largest term in this
expansion is
\begin{align}
  F_1[n(\rr)] &= \tfrac12 \iint \!\!
  g^{(2)}_{HS}(\rr_1,\rr_2)n(\rr_1)n(\rr_2)\Phi(|\rr_1-\rr_2|)
  d\rr_1d\rr_2
  \label{eq:mean-field}
\end{align}
where $g^{(2)}_{HS}(\rr_1,\rr_2)$ is the pair distribution function of
the hard-sphere reference fluid, and $\Phi(r)$ is the pair potential.
Formally, this requires the pair distribution function as a functional
of the density $n(\rr)$.  In Section~\ref{sec:gV}, we introduce
existing theoretical approaches for computing
$g^{(2)}_{HS}(\rr_1,\rr_2)$ given the external potential felt by the
hard spheres.  In Section~\ref{sec:gn}, we introduce existing
approximations for the hard-sphere pair distribution that are
expressed as a functional of the density distribution $n(\rr)$, which
is a form that is more directly useful in the construction of
classical density functionals---which are themselves expressed as a
functional of the density.

In this paper, we introduce a new contact value approach
(CVA) to approximating the hard-sphere pair distribution function
which is suitable for use in the creation of classical density
functionals based on thermodynamic perturbation theory. The resulting
function is based on a fit to the radial distribution function that is
separable in a way that enables efficient evaluation of the
integral in Eq.~\ref{eq:mean-field}.

\section{Pair distribution from the external potential}\label{sec:gV}

Given the external potential $V(\rr)$ felt by a hard-sphere fluid,
there are several approaches that have been used to compute the pair
distribution function.  We review these approaches here.  The
classic (and earliest) approach for computing the pair distribution
function given the external potential is Percus' trick of treating one
sphere as an additional contribution to the external potential, and to
find the pair distribution function from the resultant equilibrium
density~\cite{hansen2006theory}.  This elegant approach lends itself
to computation \emph{using} DFT, and can be used to compute and plot the pair
distribution function, but requires a full free-energy minimization
\emph{for each position} $\rr_1$ in $g^{(2)}(\rr_1,\rr_2)$, and hence
would be prohibitively expensive as a tool in constructing a free
energy functional.

The canonical inhomogeneous configuration for the hard-sphere fluid is
the system consisting of a hard sphere at a hard wall.  In 1986,
Plischke and Henderson solved the pair distribution function of this
system using integral equation theory under the Percus-Yevick
approximation~\cite{plischke1986pair}.
Lado recently introduced a new and more efficient algorithm for
implementing integral equation theory for inhomogeneous fluids, which
computes $g^{(2)}(\rr_1,\rr_2)$~\cite{lado2009efficient}.  While this
approach is two orders of magnitude more efficient than previous
implementations, it remains a computationally expensive approach, and
unsuitable for repeated evaluation within a free-energy minimization
as required by DFT.

Another inhomogeneous configuration that is of interest is the
test-particle configuration, in which one hard sphere is fixed.  Where
the hard-wall is a surface with no curvature, the test-particle
configuration has a surface with curvature at the molecular length
scale.  In this case, the density gives the radial distribution
function---this is just Percus' trick---and the pair distribution
function of this inhomogeneous test-particle system gives the triplet
distribution function of the homogeneous fluid.  The triplet
distribution function of the homogeneous fluid has been computed by
Gonz\'alez \emph{et al.} using the test-particle approach with
\emph{two} spheres fixed~\cite{gonzalez1999test}.

\section{Pair distribution from the density}\label{sec:gn}

The alternative to specifying the external potential is to specify the
density distribution $n(\rr)$.  One may move between these
representations by either computing the external potential
corresponding to a given density of hard spheres by taking a
functional derivative of the hard-sphere free energy functional, or by
minimizing the free energy given an external potential.  However, in
general it is simplest to use an approach that makes use of the
natural variables, which in the case of classical density functional
theory is the density.

The most direct and rigorous approach to find the pair distribution
function given the density is to take a second functional derivative
of the hard-sphere free energy to find the direct correlation
function.  One can then solve the Ornstein-Zernike equation
numerically to find the pair distribution function.  This approach was
used by G{\"o}tzelmann \emph{et al.} to solve for the pair
distribution function near a hard wall using an early hard-sphere free
energy functional~\cite{gotzelmann1996structure}.  While this approach
is rigorous, solving the inhomogeneous Ornstein-Zernike equation
remains computationally challenging, although more efficient
approximate algorithms have been developed~\cite{paul2003variational}.
This approach, while appealing, remains unsuitable for use in the
construction of a classical density functional due to its significant
computational cost.

In addition to the above exact approach, there are a number of
analytic approximations for the inhomogeneous pair distribution
function, which extend the radial distribution function to
inhomogeneous scenarios.  These approximations differ both in what
density to use when evaluating the radial distribution function
$g(r;n)$, and in how to combine the radial distribution function
evaluated at these densities~\cite{toxvaerd1973statistical}.

Early approximations to the pair distribution function used the
density at one or two positions to determine the pair distribution
function.  There are three common approaches:
\begin{align}
  g^{(2)}(\rr_1,\rr_2) &\approx
  g\left(r_{12}; n\left(\frac{\rr_1+\rr_2}{2}\right)\right)
     & \text{midpoint} \\
  g^{(2)}(\rr_1,\rr_2) &\approx
  g\left(r_{12}; \frac{n(\rr_1)+n(\rr_2)}{2}\right)
     & \text{mean density} \\
  g^{(2)}(\rr_1,\rr_2) &\approx
  \frac{g(r_{12};n(\rr_1))+g(r_{12};n(\rr_2))}{2}
     & \text{mean function}\label{eq:mean-function}
\end{align}
These approaches have been successfully and widely used in treating
the surface tension of simple fluids~\cite{pressing2003surface,
  salter2008statistical, bongiorno1975modified,
  toxvaerd1976hydrostatic, kalos1977structure, carey2008gradient,
  osborn1980monotonic, mccoy1981comparison, barrett2006some}.  The
mean density approximation has also been quoted (as a goal) by recent
papers that proceed to make further
approximations~\cite{gloor2007prediction, gross2009density}.  However,
these approximations fail dramatically when applied to strongly
inhomogeneous systems such as a dense fluid at a solid surface.  Such
systems exhibit a strongly oscillatory density distribution, with
density peaks that can have local packing fractions greater than
unity, which cannot occur in the bulk reference system that defines
$g(r; n)$.  The above papers restrict themselves to the liquid-vapor
interface, which does not exhibit this pathology.

Non-pathological approaches use an average of the density over some
volume. Fischer and Methfessel introduce the
approximation~\cite{fischer1980born,harris1987comment}:
\begin{align}
  g^{(2)}(\rr_1,\rr_2) \approx g\left(r_{12}; n_3\left(\tfrac12
  (\rr_1+\rr_2)\right)\right)
  \label{eq:fischer}
\end{align}
where $n_3$ is an integral of the density over a spherical volume that
is now used as one of the fundamental measures in Fundamental
Measure Theory~\cite{rosenfeld1989free}:
\begin{align}
  n_3(\rr) = \int n(\rr')\Theta(\tfrac12 \sigma - |\rr-\rr'|) d\rr'
\end{align}
Equation~\ref{eq:fischer} is computationally awkward, because it
treats as special the midpoint $\tfrac12(\rr_1+\rr_2)$.  Moreover, the
approach of Fischer and Methfessel is intended to approximate the pair
distribution function only at contact, when the distance between
$\rr_1$ and $\rr_2$ is the hard-sphere diameter.
Tang \emph{et al.} employed an approximation for the pair distribution
function that is similar to that of Fischer and Methfessel, but with a
self-consistent weighted density computed with a weighting function that
is itself dependent on the weighted density~\cite{tang1991density}.
This weighted density was computed using the hard-sphere weighted
density of Tarazona, which was developed using the direct correlation
function of the homogeneous hard-sphere fluid~\cite{tarazona1985free}.

Sokolowski and Fischer addressed the shortcomings of the theory of
Fischer and Methfessel by modifying this approach to use density
averages centered on the two points $\rr_1$ and $\rr_2$:
\begin{align}
  g^{(2)}(\rr_1,\rr_2) \approx g\left(r_{12};
  \tfrac12(\bar{n}(\rr_1)+\bar{n}(\rr_2))\right)
  \label{eq:sokolowski}
\end{align}
where their averaged density $\bar{n}(\rr)$ given by
\begin{align}
  \bar{n}(\rr) \equiv \frac{3}{4\pi (0.8\sigma)^3}\int n(\rr')\Theta(0.8\sigma - |\rr-\rr'|) d\rr'
\end{align}
is the density averaged over a sphere with diameter
$0.8\sigma$~\cite{sokolowski1992role}.  The value 0.8 in this formula
was arrived at by fitting to Monte Carlo simulation.  Although
Eq.~\ref{eq:sokolowski} has the advantage of only involving
density averages at the points at which the pair distribution function
is desired, it remains sufficiently computationally cumbersome that it
has only been used in two papers studying the one-dimensional liquid vapor
interface~\cite{wadewitz2000application, winkelmann2001liquid}.
Because it cannot be written as a single-site convolution, this
approach is particularly computationally demanding when applied to
systems featuring inhomogeneity in more than one dimension.

In a previous paper~\cite{schulte2012using}, we introduce a
functional that gives a good approximation for the pair distribution
function averaged over positions $\rr_2$ that are in contact with
$\rr_1$, defined as:
\begin{align}
  g_\sigma(\rr_1) &\equiv \frac{ \int g^{(2)}(\rr_1,\rr_2) \delta(\sigma -|\rr_1-\rr_2|)n(\rr_2)
    d\rr_2 }{ \tilde{n}(\rr_1)  }
\end{align}
where the weighted density $\tilde{n}(\rr_1)$ is defined by:
\begin{align}
  \tilde{n}(\rr) &\equiv \int n(\rr') \delta(\sigma -|\rr - \rr'|)d\rr'.
\end{align}
In \cite{schulte2012using} we use the contact-value theorem to derive the exact formula:
\begin{align}
  g_\sigma(\rr)
  &= \frac12 \frac{1}{k_BT n(\rr) \tilde{n}(\rr)} \frac{\delta
    F_{HS}}{\delta \sigma(\mathbf{r})} \label{eq:gsigma-exact}
\end{align}
where $\sigma(\rr)$ is the diameter of hard spheres located at
position $\rr$, and $F_{HS}$ is the Helmholtz free energy of the
hard-sphere fluid.  The functional derivative of the free energy with
respect to the hard-sphere diameter in Eq.~\ref{eq:gsigma-exact} requires
that we be able to evaluate the change in free energy resulting from a
change in the diameter of specifically the hard spheres located at
position $\rr$.  This somewhat unusual construction is mathematically
straightforward within Fundamental Measure Theory
(FMT)~\cite{rosenfeld1989free}.  We employ the White Bear variation of
the FMT free energy functional~\cite{roth2002whitebear}, which
provides an excellent approximation for this averaged value of the
pair distribution function at contact for a variety of interfaces, and
over a wide range of densities.



\section{Contact value approach}
In the approaches for the pair distribution function
mentioned above, the radial distribution function used in the
approximation was dependent upon the density averaged over some
volume.  We seek to achieve greater accuracy by making use of a
function dependent upon our averaged $g_{\sigma}(\rr)$ discussed
above, which holds more information about an inhomogeneous system than
does a simple convolution of the density.
We construct the CVA with the average of two radial distribution
functions, evaluated at the distance between the two points, that are
themselves functions of the averaged pair distribution function at contact
$g_{\sigma}(\rr)$ evaluated at the two points:
\begin{align}
  g^{(2)}(\rr_1,\rr_2) = \frac{g(r_{12}; g_\sigma(\rr_1)) +
    g(r_{12}; g_\sigma(\rr_2))}{2}. \label{eq:g2-our-mean}
\end{align}
This Contact Value Approach for $g^{(2)}(\rr_1,\rr_2)$ is
constructed to reproduce the exact value for the integral:
\begin{align}
  F_1^{\text{contact}} &= \tfrac12 \iint
  g^{(2)}_{HS}(\rr_1,\rr_2)n(\rr_1)n(\rr_2)\delta(|\rr_1-\rr_2|-\sigma)
  d\rr_1d\rr_2
  \label{eq:mean-field-contact}
\end{align}
which is the mean-field correction to the free energy (see
Eq.~\ref{eq:mean-field}) for a purely contact interaction.

The CVA requires the radial distribution function expressed as a
function of $r$ and $g_\sigma$.
%
%
We construct a functional form for $g(r, g_\sigma)$ that allows for
improved computational efficiency.  We introduce the general form that
allows for this efficiency in Section~\ref{sec:efficient}, and we
detail our specific approximation for $g(r,g_\sigma)$ that uses this
general form in Section~\ref{sec:separable-fit}.
%

\newcommand\maxrfit{4}
\newcommand\maxerr{0.2}
\newcommand\etamaxerr{0.45}
\newcommand\rmaxerr{3.7}
\newcommand\chisq{2.5}
\newcommand\kappatable{
  \left(
  \begin{array}{c d{3} d{3} d{3} d{3}}
    -1.754 & 0.027 & 0.838 & -0.178 \\
    -2.243 & 4.403 & -2.48 & 0.363 \\
    0.207 & 0.712 & -1.952 & 1.046 \\
    -0.002 & -0.164 & 0.324 & -0.162 \\
  \end{array}
  \right)
}
\newcommand\alphaval{-0.002}

\section{Making the CVA efficient}
\label{sec:efficient}
The existing approaches to approximating the pair distribution
function outlined in Section~\ref{sec:gn} have not been widely used in
the construction of density functionals based on thermodynamic
perturbation theory, largely due to their computational complexity.
While our CVA provides only an incremental improvement in accuracy,
its construction enables significant gains in computational
efficiency, allowing for practical application in density functionals.
We achieve this gain by developing a \emph{separable} fit to the radial
distribution function of the hard-sphere fluid (see
Section~\ref{sec:separable-fit} for details).  This separable fit is
of the form
\begin{align}
  g(r; g_\sigma) &= \sum_{i} a_i(r) b_i(g_\sigma)
\end{align}
where the notable aspect is that the radial distribution function is
written as a sum of terms that are each a simple product of a function
of radius with a function of $g_\sigma$.  This enables us to write
integrals---such as Eq.~\ref{eq:mean-field}---that are linear in the
pair distribution function as a summation of fixed-kernel
convolutions, which may be efficiently computed using Fast Fourier
Transforms (FFTs).

\newcommand\Vcell{V_{\textit{cell}}}
\newcommand\Vinteraction{V_{\Phi}}

Computation of the free energy correction from Eq.~\ref{eq:mean-field}
for a periodic system by direct integration requires a nested
integration over the volume of the system~$\Vcell$, and the volume
over which the interaction is nonzero~$\Vinteraction$.  Thus the cost
of computation scales as
$\mathcal{O}\left(\frac{\Vcell\Vinteraction}{\Delta V^2}\right)$ where
$\Delta V$ is the volume resolution of the computational grid.  Direct
integration is the most efficient algorithm when using the existing
functionals for $g^{(2)}(\rr_1,\rr_2)$ described in
Section~\ref{sec:gn}.  The one exception is the ``mean-function''
approximation (Eq.~\ref{eq:mean-function}), which could in principle
be made more efficient using the same technique we describe here.
Because the CVA allows the integral in Eq.~\ref{eq:mean-field} to be
written as a sum of fixed-kernel convolutions, it can be computed
without a nested integral, at the cost of performing a few FFTs.  This
approach scales as $\mathcal{O}\left(\frac{\Vcell}{\Delta
  V}\log\frac{\Vcell}{\Delta V}\right)$, as do most widely used DFT
functionals such as FMT~\cite{rosenfeld1989free, roth2002whitebear}.
With this scaling, when examining systems with long interaction
distances or high resolution---which is often necessary when working
with hard-sphere functionals---the CVA has the potential to be far
more efficient than existing methods.

\begin{figure}
  \centering
  \includegraphics[width=\columnwidth]{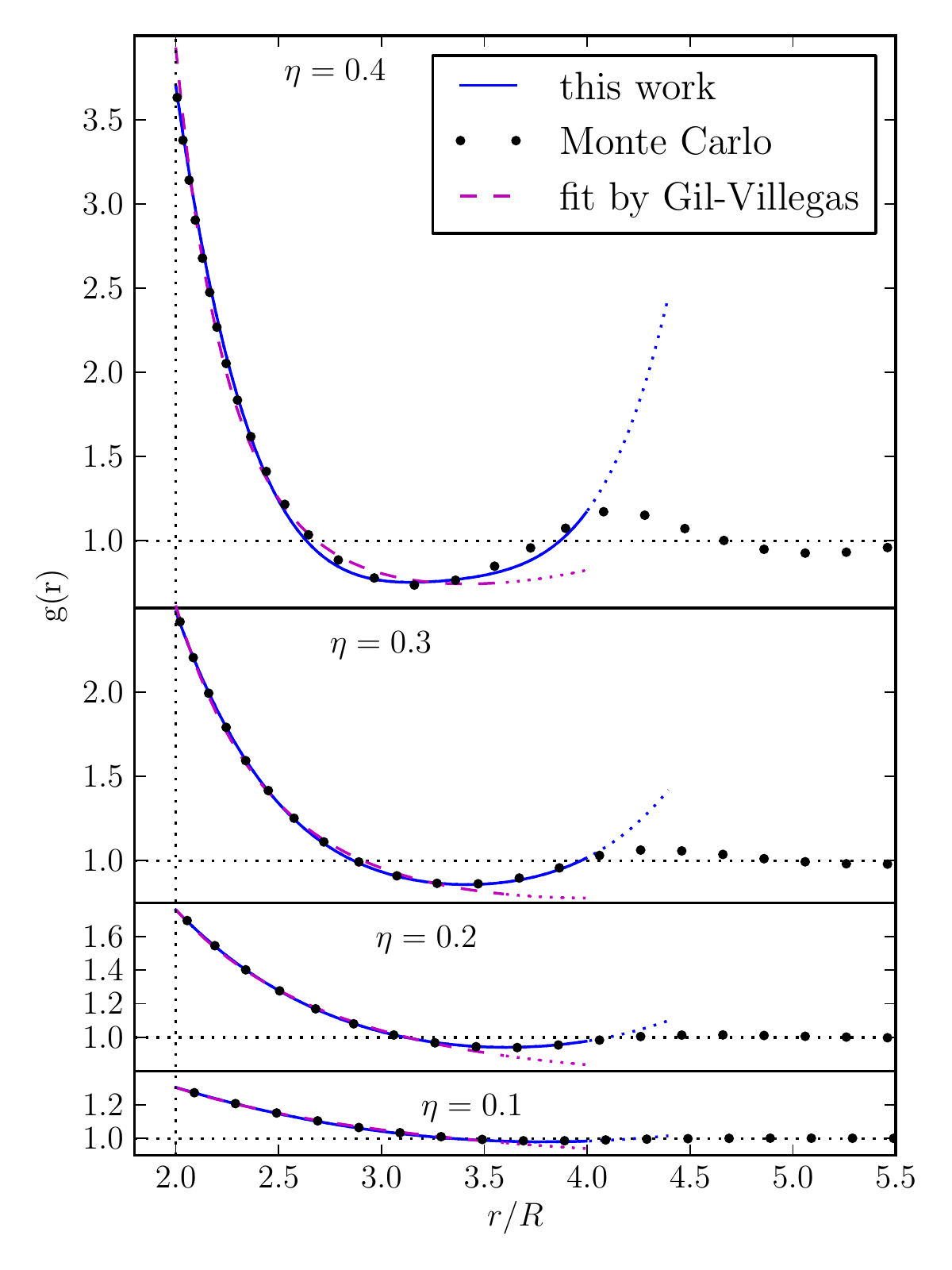}
  \caption{Plot of the hard-sphere radial distribution function of the
    homogeneous fluid at several values for packing fraction $\eta$. The
    blue lines show our separable fit, the black dots show the true
    radial distribution function $g(r)$ as found from Monte Carlo
    simulation, and the dashed lines are results of the
    Gil-Villegas fit~\cite{gil1997statistical}.  The dotted extension
    of each fitted curve indicates the value of the function outside
    of the fitted region.  }\label{fig:radial-distribution}
\end{figure}

To see how we obtain this improved scaling, we examine the
lowest-order correction in TPT, given by Eq.~\ref{eq:mean-field}.  The
two terms that are averaged in Eq.~\ref{eq:g2-our-mean} give equal
contributions to the integral
\begin{align}
  F_1^{\textit{CVA}} &= \tfrac12 \!\! \iint \!\!
  g(r_{12};g_\sigma(\rr_2))n(\rr_1)n(\rr_2)\Phi(|\rr_1-\rr_2|)
  d\rr_1d\rr_2.
\end{align}
When we introduce the separable form for $g(r_{12};g_\sigma)$ we can
further simplify this integral as
\begin{align}
  F_1^{\textit{CVA}} \! = \!
  \sum_i \!\tfrac12 \!\! \int \!\! n(\rr_1) \!\!
                            \int \!\!\!
                            a_i(r_{12})\Phi(r_{12})
                            b_i(
g_\sigma(\rr_2))n(\rr_2)
  d\rr_2d\rr_1
\end{align}
where the functional is written as a summation of integrals of simple
convolutions in three dimensions.  Thus, each of these integrals may
be computed in $\mathcal{O}(N\log N)$ time, where $N$ is the number of
grid points in the computational cell.  This is the same scaling as is
required to compute the fundamental measures such as $n_3$ which are
used in FMT.

\section{A separable fit for the radial distribution function}\label{sec:separable-fit}

Having settled on the basic structure of our function, we further
refine it by performing a separable fit to the radial distribution
function from Monte Carlo simulation.  We focus our fit on
the range of distances $r_{12} \le \maxrfit R$.  This range is
relevant to the widely used \cite{chapman1989saft,
  muller2001molecular, tan2008recent} Statistical
Associating Fluid Theory of Variable Range (SAFT-VR) free energy with
square-well dispersive attraction developed by Gil-Villegas \emph{et
  al.}~\cite{gil1997statistical}.
Although we consider this range of radii particularly interesting,
this is not a fundamental limit of the approach, as one could readily
extend the fit to larger radii by including additional fitting
parameters.
For comparison, in Fig.~\ref{fig:radial-distribution} we plot our fit,
Monte-Carlo data, and the radial distribution function of Gil-Villegas
\emph{et al.}, which we have extracted from their approximation for
the first term in the dispersion free energy given by
Eq.~\ref{eq:mean-field}.

For ease of implementation and future extension to larger radii, we
fit the radial distribution function using a fourth-order polynomial.  We constrain
our functional form such that $g(r; g_\sigma)$ reduces to $g_\sigma$
at contact and approaches $g(r)=1$ in the low-density limit.
Incorporating these constraints we have the functional form
\begin{align}
  g(r;g_\sigma) &=
  g_\sigma + \sum_{i=1}^{4} \sum_{j=1}^{4} \kappa_{ij} (g_\sigma - 1)^i
  \left(\tfrac{r}{\sigma}-1\right)^j,
  \label{eq:fit-form}
\end{align}
where the matrix $\kappa_{ij}$ is determined from a least-squares
fit to Monte Carlo data for the radial distribution function, over the
range $2R \le r \le \maxrfit R$, and for packing fractions $\eta \le
0.45$.  The resulting parameters are displayed in
Table~\ref{tab:kappa}.  The maximum error in $g(r)$ within this
range is \maxerr, which occurs at $\eta = \etamaxerr$ and $r =
\rmaxerr R$.  Fig.~\ref{fig:radial-distribution} displays
our approximation at just under half of the densities that were
included in the fit.

\begin{table}
  \begin{align*}
    \kappa &= \kappatable
  \end{align*}
  \caption{The fitted $\kappa_{ij}$ matrix.
  }\label{tab:kappa}
\end{table}

\newcommand{\plotcomp}[1]{The top halves of
  these figures show the results of Monte Carlo simulations, while the
  bottom halves show the CVA, truncated beyond the range of the fit.
  On the right are plots of #1 on the
  paths illustrated in the figures to the left.  These plots compare
  the CVA (blue solid line), Monte Carlo results (black circles), the
  results of Sokolowski and Fischer (red dashed
  line)~\cite{sokolowski1992role}, and those of Fischer and Methfessel
  (green dot-dashed line)~\cite{fischer1980born}.  The latter is only
  plotted at contact, where it is defined}

\begin{figure*}
  \begin{subfigure}{\textwidth}
    \includegraphics[width=\linewidth]{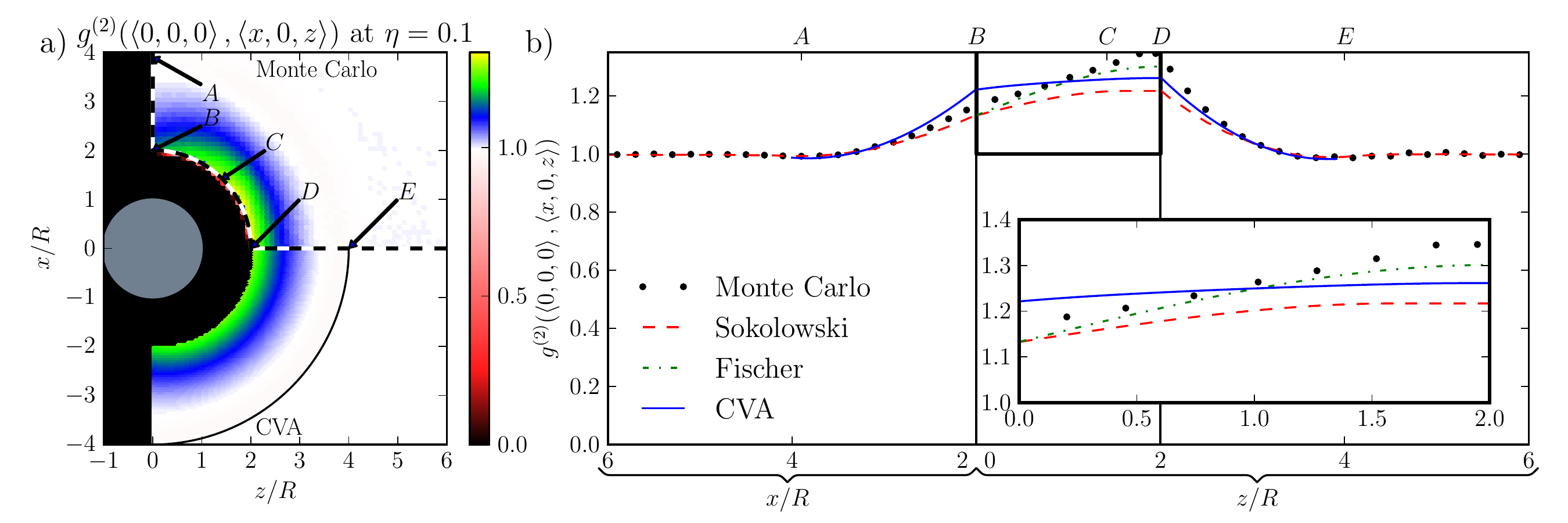}
    \vspace{-0.6cm}
  \end{subfigure}
  \begin{subfigure}{\textwidth}
    \includegraphics[width=\linewidth]{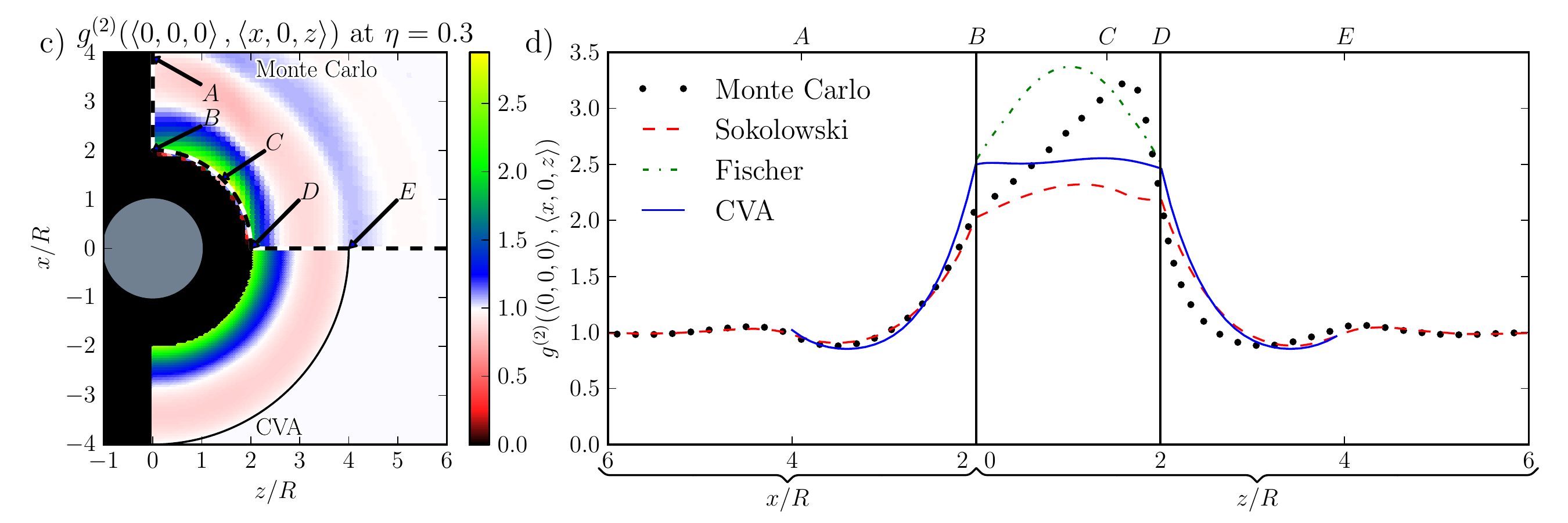}
    \vspace{-0.6cm}
  \end{subfigure}
  \caption{The pair distribution function near a hard wall, with
    packing fractions of 0.1 and 0.3 and $\rr_1$ in contact with the
    hard wall.  On the left are 2D plots of $g^{(2)}(\rr_1,\rr_2)$ as
    $\rr_2$ varies. \plotcomp{$g^{(2)}(\rr_1,\rr_2)$}.}
  \label{fig:pair-distribution}
\end{figure*}

\section{Results}

\subsection{Pair distribution function}


We begin by examining the pair distribution function near a hard wall,
with a focus on the case where one of the two spheres is in contact
with the hard wall.  Figures~\ref{fig:pair-distribution}a
and~\ref{fig:pair-distribution}c compare the results of the CVA with
Monte Carlo simulations at packing fractions of 0.1 and 0.3
respectively. We see reasonable agreement at the lower density, with a flatter angular
dependence when the two spheres are in contact.  At the higher
density, we see significant structure developing in the simulation
results that is not reflected in our approximation.

\begin{figure*}
  \includegraphics[width=\textwidth]{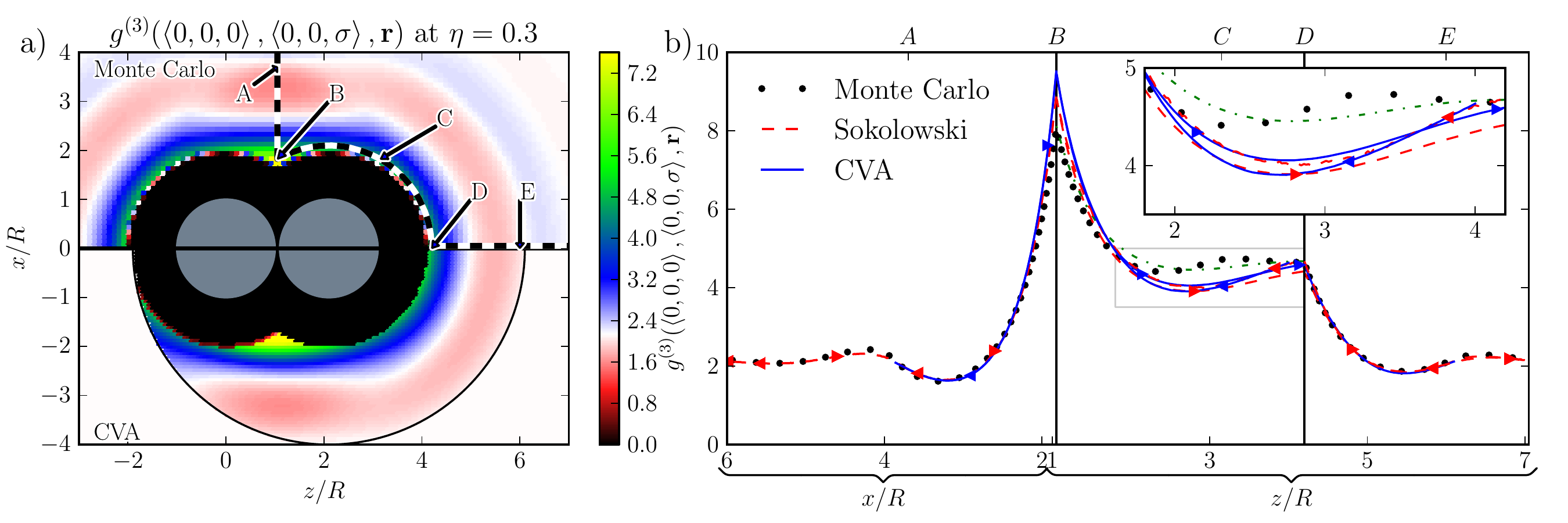}
  \vspace{-0.7cm}
\\
  \includegraphics[width=\textwidth]{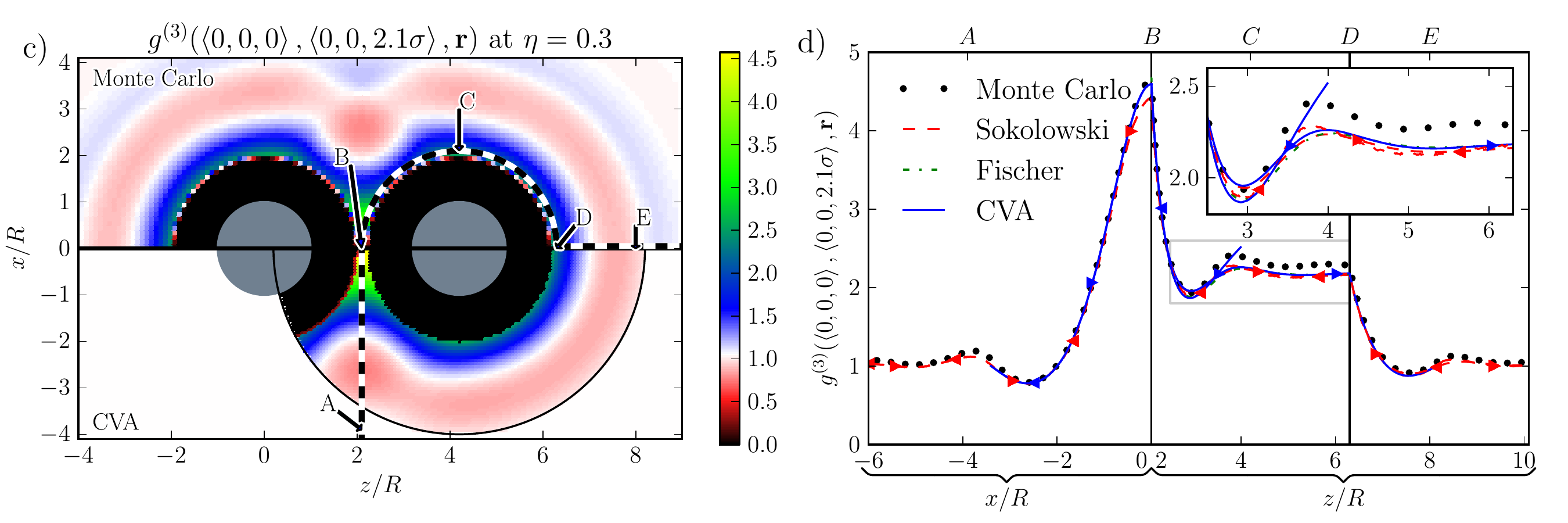}
  \vspace{-0.7cm}
  \caption{The triplet distribution function
    $g^{(3)}(\rr_1,\rr_2,\rr_3)$ at packing fraction 0.3, plotted when
    $\rr_1$ and $\rr_2$ are in contact (a,b) and when $\rr_1$ and
    $\rr_2$ are separated by a distance $2.1\sigma$ (c,d). On the left
    are 2D plots of $g^{(3)}(\rr_1,\rr_2,\rr_3)$ as $\rr_3$
    varies. 
    The top halves of these figures show the results of Monte Carlo
    simulations, while the bottom halves show the CVA, truncated
    beyond the range of the fit.  On the right
    are plots of $g^{(3)}(\rr_1,\rr_2,\rr_3)$ on the paths illustrated
    in the figures to the left.
    We also plot these curves along a left-right mirror image of this
    path.  The data for the right-hand paths (as shown in the 2D
    images) are marked with right-pointing triangles, while the
    left-hand paths are marked with left-pointing triangles.
%
  }
  \label{fig:triplet-contact-distribution}
\end{figure*}

Figures~\ref{fig:pair-distribution}b and~\ref{fig:pair-distribution}d
show the pair distribution function as plotted along paths illustrated
in Figures~\ref{fig:pair-distribution}a
and~\ref{fig:pair-distribution}c.  These plots compare the CVA with
Monte Carlo results, as well as the approximations of Sokolowski and
Fischer~\cite{sokolowski1992role} and of Fischer and
Methfessel~\cite{fischer1980born} at the same packing fractions of 0.1
and 0.3.  The approach of Fischer and Methfessel is only defined when
the two spheres are in contact, and is therefore only plotted on that
segment of the path.  As an input to the previous approximations we
use the hard sphere radial distribution function found with Monte
Carlo simulation, interpolated as necessary. We find that both
previous approximations to the pair distribution function predict
stronger angular dependence of the pair distribution function at
contact than this work.  The previous approximations each have a
systematic error at contact---either too high or too low.  In
contrast, our errors at contact have a tendency to cancel when used in
a perturbation expansion.  At higher densities, the approximation of
Fischer and Methfessel requires evaluating the radial distribution
function at densities significantly higher than the freezing density,
which poses numerical difficulties when using the radial distribution
function from simulation.  When the two points $\rr_1$ and $\rr_2$ are
both more than a radius away from contact, we find that any of these
approaches gives a reasonable prediction.

\subsection{Triplet distribution function}

Just as the radial distribution function of a homogeneous fluid may be
computed from the density of an inhomogeneous one using Percus'
test-particle trick, the triplet distribution function of a
homogeneous system can be computed using an approximation of the pair
distribution for an inhomogeneous fluid, such as we have
developed. The triplet distribution function of a homogeneous fluid
with density $n$ is given by:
\begin{multline}
    g^{(3)}(\rr_1,\rr_2,\rr_3) =\\
    \frac{n_{\textrm{TP}(\rr_1)}(\rr_2)
      n_{\textrm{TP}(\rr_1)}(\rr_3)}{n^2}
    g^{(2)}_{\textrm{TP}(\rr_1)}(\rr_2,\rr_3)
\end{multline}
where the $\textrm{TP}(\rr_1)$ subscript indicates quantities computed for
the inhomogeneous density configuration in which one sphere (the
``test particle'') is fixed
at position $\rr_1$.  This method treats one of the three
positions---the location of the test particle---differently from the
other two, which means that a poor approximation to the pair distribution
function may break the symmetry between $\rr_1$ and $\rr_2$ which is
present in the true triplet distribution function.

Figures~\ref{fig:triplet-contact-distribution}a
and~\ref{fig:triplet-contact-distribution}c compare the triplet
distribution function at a packing fraction of 0.3 computed using the
CVA with results from Monte Carlo simulations. In
Figure~\ref{fig:triplet-contact-distribution}a the spheres at $\rr_1$
and $\rr_2$ are in contact; in
Figure~\ref{fig:triplet-contact-distribution}c they are spaced so that
a third sphere can just fit between them; and in both figures $\rr_3$
is varied. The test-particle position for the CVA in each case is
$\rr_1$, which is on the left-hand side of the figure. As before, we
see reasonable agreement with simulation. Also, the Monte Carlo
results have the expected left-right symmetry, while the CVA has a
small asymmetry introduced with the test particle due to errors in the
pair distribution function.

Figures~\ref{fig:triplet-contact-distribution}b
and~\ref{fig:triplet-contact-distribution}d show the triplet
distribution function as plotted along the paths illustrated in
Figures~\ref{fig:triplet-contact-distribution}a
and~\ref{fig:triplet-contact-distribution}c.  We also plot the
results along a left-right mirror image path, corresponding to
swapping $\rr_1$ and $\rr_2$. The two mirror-image paths are
distinguished by arrows (triangles) along the curves, with right-facing arrows
indicating the paths shown in
Figures~\ref{fig:triplet-contact-distribution}a and
\ref{fig:triplet-contact-distribution}c, and left-facing arrows
indicating the mirror image path.  As the work of
Fischer and Methfessel is only defined when $\rr_2$ and $\rr_3$ are in
contact, we only plot it along the
central portion of the path, which is in contact with $\rr_2$, and arrows
are omitted.
All methods tested perform similarly over their range of validity.

\begin{figure}
  \begin{subfigure}{1.0\columnwidth}
    \includegraphics[width=\columnwidth]{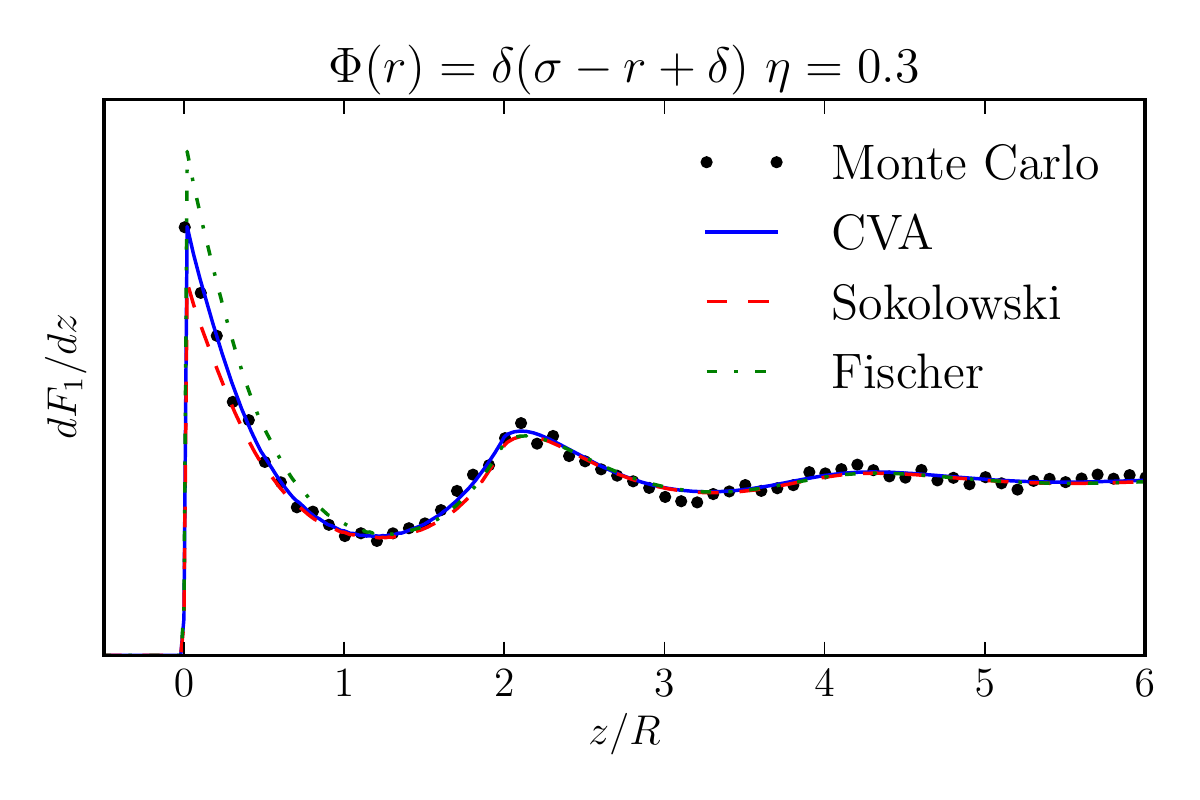}
    \vspace{-0.8cm}
    \caption{Sticky hard-sphere fluid}\label{fig:dadz-delta}
  \end{subfigure}
  \begin{subfigure}{1.0\columnwidth}
    \includegraphics[width=\columnwidth]{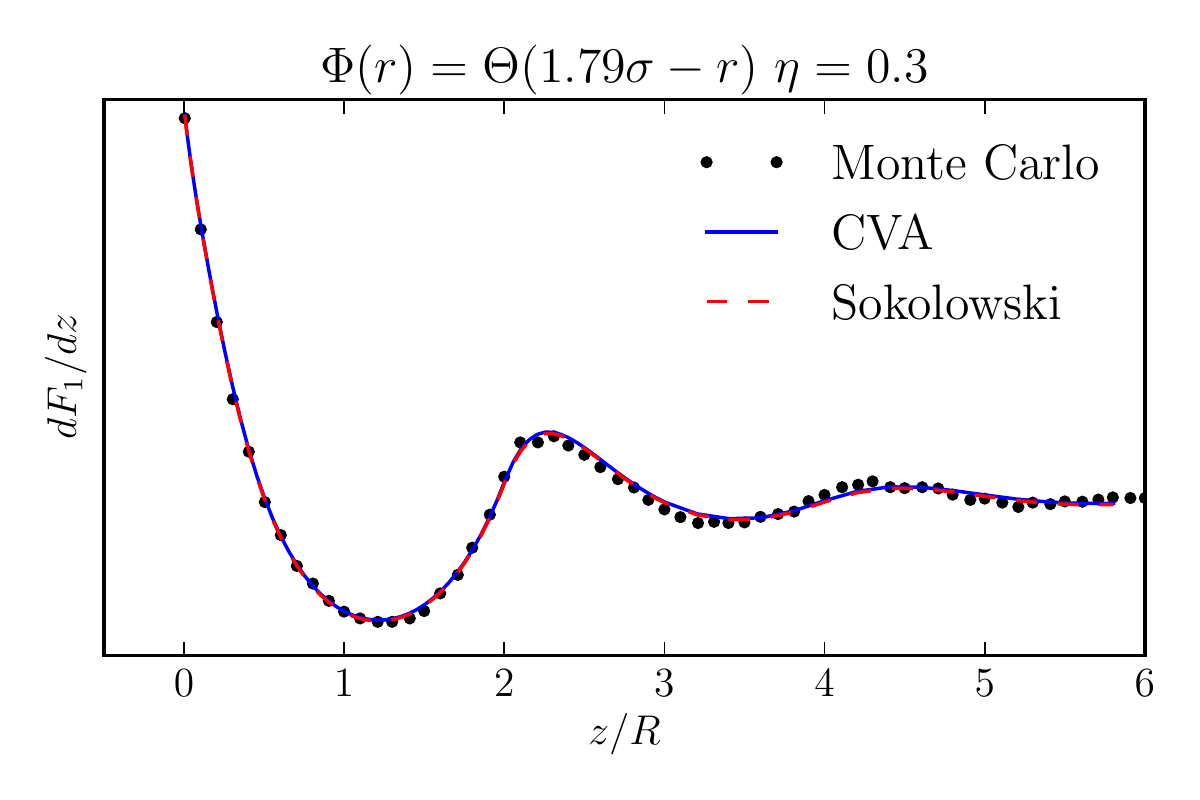}
    \vspace{-0.8cm}
    \caption{Hard-core square well fluid}\label{fig:dadz-square-well}
  \end{subfigure}
  \caption{Plot of $\frac{dF_1}{dz}$ near a hard wall, with arbitrary
    vertical scale.  (a) shows a
    sticky hard-sphere fluid defined by a pair potential
    $\delta(\sigma-r+\delta)$ where $\sigma$ is the hard-sphere
    diameter, and $\delta$ is an infinitesimal distance; and (b) shows a
    square well fluid defined by a pair potential $\Theta(1.79
    \sigma-r)$.
  }
  \label{fig:dadz}
\end{figure}

\section{Accuracy in thermodynamic perturbation theory}

A particularly relevant quantitative test of a pair distribution
function is how well it predicts the interaction energy due to a pair
potential.  To this end, we have computed the error in the first term
in a high-temperature perturbation expansion $F_1$
for two typical pair potentials.  In order to focus on effects at the
interface, we have defined a position-dependent pair interaction
energy as
\begin{align}
  \frac{dF_1}{dz} &=
  \tfrac12 \int g^{(2)}_{HS}(\rr,\rr')n(\rr)n(\rr')\Phi(|\rr-\rr'|)
  d\rr'\, dxdy\label{eq:da1}
\end{align}
which gives the contribution to the mean-field free energy due to
molecules located in a plane of fixed $z$.

We plot this quantity for two representative pair potentials near a
hard wall in Fig.~\ref{fig:dadz}.  We have chosen to illustrate a
delta-function interaction at contact (i.e. ``sticky hard spheres''),
and a hard-core square-well fluid, with the length-scale of
interaction taken from the optimal SAFT model for water found by Clark
\emph{et al.}~\cite{clark2006developing}.  These pair potentials
represent both a very short-range interaction and a medium-range
interaction.

Figure~\ref{fig:dadz-delta} shows the results for the sticky
hard-sphere fluid.  The CVA is constructed to produce this result
exactly, provided the averaged pair distribution function at contact
from Ref.~\citenum{schulte2012using} is exact.  As expected, we see
excellent agreement with the Monte Carlo simulation results, while the
approximations of Fischer and Sokolowski each show deviations near the
interface.  Figure~\ref{fig:dadz-square-well} shows the same curve
from Eq.~\ref{eq:da1} for the square-well fluid.  In this case both
the CVA and Sokolowski's approximation give excellent agreement with
simulation.

\section{Conclusion}

We have introduced and tested the contact value approach for the pair
distribution function $g^{(2)}(\rr_1,\rr_2)$ of the inhomogeneous
hard-sphere fluid.  The pair distribution function plays a key role in
thermodynamic perturbation theory, which is widely used in the
construction of classical density functionals.  The CVA---unlike
existing approximations---is suitable for use in classical density
functionals based on perturbation theory, as it may be efficiently
computed using exclusively fixed-kernel convolutions.  We have tested
this function at a hard wall and near a single fixed hard sphere, and
find that it gives excellent agreement with simulation.  Tests of the
pair distribution function in integrals that arise in thermodynamic
perturbation theory suggest that the CVA is accurate for attractions
up to the distance to which the radial distribution function is fit,
and is a significant improvement over existing approximations near
contact.  But most importantly, the computational cost of using the
CVA in a classical density functional scales much more favorably than
existing methods in high resolution computations.

\bibliography{paper}

\end{document}